# Concentration Polarization and Metal Dendrite Initiation in Isolated Electrolyte Microchannels

Youngju Lee[a], Bingyuan Ma[a], Peng Bai[a,b,*]

[a] Department of Energy, Environmental and Chemical Engineering, Washington University in St. Louis, 1 Brooking Dr, St. Louis, MO 63130, USA.

[b] Institute of Materials Science and Engineering, Washington University in St. Louis, 1 Brooking Dr, St. Louis, MO 63130, USA.

*Correspondence to: pbai@wustl.edu

**Abstract**

Lithium metal penetrations through the liquid-electrolyte-wetted porous separator and solid electrolytes are a major safety concern of next-generation rechargeable metal batteries. The penetrations were frequently discovered to occur through only a few isolated channels, as revealed by “black spots” on both sides of the separator or electrolyte, which manifest a highly localized ionic flux or current density. Predictions of the penetration time have been infeasible due to the hidden and unclear dynamics in these penetration channels. Here, using the glass capillary cells, we investigate for the first time the unexpectedly sensitive influence of channel geometry on the concentration polarization and dendrite initiation processes. The characteristic time for the complete depletion of salt concentration on the surface of the advancing electrode, i.e. Sand’s time, exhibits a nonlinear dependence on the curvature of the channel walls along the axial direction. While a positively deviated Sand’s time scaling exponent can be used to infer a converging penetration area through the electrolyte, a negatively deviated scaling exponent suggests that diffusion limitation can be avoided in expanding channels, such that the fast-advancing tip-growing dendrites will not be initiated. The safety design of rechargeable metal batteries will benefit from considering the true local current densities and the conduction structures.

# 1 Introduction

Owing to the high specific capacity and low redox potential, Li metal anodes are being developed and optimized for the next-generation rechargeable metal batteries working at higher areal capacities and higher power densities.[1–4] In these harsher working conditions, batteries are prone to develop nonuniform ionic flux in the electrolyte, leading to localized deposition of metallic Li along electrolyte channels that can cause internal shorts and low cycle life.[1,5] While promising methods of tuning the electrolyte additive,[1,6,7] surface charge of the separator,[8,9] functionalities of the artificial solid-electrolyte interphase (SEI) layers,[10–12] and solid-state electrolytes[13–15] have been proposed to control the root-growing whiskers and surface-growing clusters of lithium metal at under-limiting current densities, there exists no clear solution to block tip-growing Li dendrites at over-limiting current density in liquid electrolytes, or at the empirical critical current density (CCD) in solid electrolytes.

Taking liquid electrolytes as the model system, the limiting current density is a system-specific property determined mainly by the equilibrium concentration ($c_0$), interelectrode distance ($L$), ambipolar diffusion coefficient ($D_{\mathrm{amb}}$), and transference number of the anion ($t_{\mathrm{a}}$), via the formula[16] $J_{lim} = 2z_c c_0 F D_{amb} (t_a L)^{-1}$. In practical cells with a very thin separator or electrolyte layer between the two electrodes, the theoretical limiting current density can be ~100 mA cm$^{-2}$. Practical current densities (< 10 mA cm$^{-2}$) will not trigger diffusion limitation, but only if the applied total current were uniformly distributed over the entire geometrical area of the electrolyte or separator. However, recent studies of failure mechanisms in rechargeable metal batteries showed that metal penetrations were self-confined within a few isolated microchannels of the porous separator,[17–20] or highly localized paths through polymer[21–26] and ceramic solid electrolytes,[27,28] as revealed by the isolated "black spots" visible on both sides of the separator or electrolyte. The highly localized penetration spots manifest that the local current density that enabled the penetration must have been several orders of magnitude higher than the geometrically averaged current density and may even be higher than the system-specific limiting current density. Accurate understandings of the concentration polarization dynamics in these isolated electrolyte microchannels are important for strategies to avoid or stop the fast-advancing tip-growing dendrites[5] in rechargeable metal batteries. Apparently, the penetration channels were not ideally straight especially for the systems containing random-porous separators,[29–31] yet the interelectrode distances would not remain fixed due to mossy growths before Sand's time.[5,27] These irregularities of the electrolyte microchannels, while appear negligible, can cause substantial deviation of the transport dynamics from the classical understandings.[32]

Here, we use our glass capillary cells as a model system to investigate the transport dynamics in isolated electrolyte microchannels. The good agreement between *operando* experiments and mathematical simulations reveals the strong dependence of concentration polarization on the geometry of the channels, as reflected by the scaling exponent between Sand's times obtained at various overlimiting current densities. Our analytical solution to the transport dynamics provides a novel implicit Sand's time formula as a function of the applied current density. Depending on the curvature and electrode advancing velocity, deviation of the Sand's time scaling varies quite dramatically. Our results suggest that separators with expanding ionic conduction structures can delay the concentration depletion at the electrode surface, which can be utilized to avoid the fast-advancing tip-growing dendrites.

# 2 Results

## 2.1 Current Densities in the Penetration Channels

As motivation, we begin by estimating the current density focused onto the penetration "black spots" reported in the literature. Table 1 summarizes the ratio between the areas of the black spots ($A_p$) and the electrode ($A_{geo}$). The significantly small ratios lead to dramatic amplification of the true local current densities that effected the penetrations. More details are available in the Supplementary Information (Fig. S1). Even by the most conservative estimation, the current density through the penetration spots ($J_p$) appears to be three orders of magnitude higher than the original design.

**Table 1. Estimation of the penetration current density focused on the "black spots" areas.**

| **Electrolyte** | $\boldsymbol{A_p/A_{geo}}$ **(%)** | $\boldsymbol{J_p}$ **(**mA cm$^{-2}$**)** |
|---|---|---|
| Liquid | 0.16-0.7 | 70-2400 |
| Ceramic | 0.06-0.36 | 30-60 |
| Polymer | 0.005-0.16 | 200-10000 |

The geometry of electrolyte microchannels depends on the pore structures of the separator or the grains/domain structures of the solid electrolytes. It also depends on the mechanical interactions between the metal deposits and the channel walls.[33–37] Transport dynamics inside precisely controlled ideal microfluidic channels have been investigated extensively,[38,39] but that in the more practical irregular channels are to be studied. A very simple yet mild irregularity can cause significant deviation from the classical understandings,[5] which necessitates a systematic analysis of the dynamics in practice-relevant conditions.

As we previously reported, examining the system-specific limiting current density and the Sand's time, i.e. the characteristic time for an electrochemical cell to have the complete concentration depletion near the electrode surface at an overlimiting current density, provides the reliable determination of the two physical parameters that control the transport dynamics, i.e. diffusion coefficient and transference number.[5] Our glass capillary cells with the inner diameter as small as 90 μm closely resemble the characteristic geometry of the penetration channels observed in various experimental cells. Monitoring the transparent cells offers straightforward evidence of the dynamics, which otherwise are confined within channels buried inside opaque materials of the separators or solid electrolytes.

## 2.2 Sand's Time in Converging Channels

As revealed by the X-ray computer tomography (CT) image of the pulled glass capillary in Fig. 1a, the inner diameter ($r$) versus the axial position ($x$) can be fitted by a catenary curve,[40] $r(x) = a\cosh(x/a) - b$, where $a = 70.640$ mm and $b = 70.595$ mm are curve constants. When the mixed ethylene carbonate and dimethyl carbonate (EC/DMC) solvent is used, lithium metal electrodes can only be pushed into the tapering part of the pulled glass capillary to form a symmetrical cell with an interelectrode distance around 5 mm. Within this long-distance, changes of the inner diameter, from ~ 180 μm (edge) to ~ 90 μm (center), are clearly visible (Fig. 1c). When triglyme solvent is used instead, the lithium metal electrode can be further pushed toward the center part. As shown in Figure 1b, the inner wall of the capillary cell within the center 2-mm length is almost straight (Fig. 1b). We were able to fabricate long cells with both the converging and expanding parts, as well as short cells with a nearly straight inner wall. A comparative study using these three types of cells (converging, straight and expanding channels) helps to understand the deviation of transport dynamics from the classical understandings, due to the geometric effect.

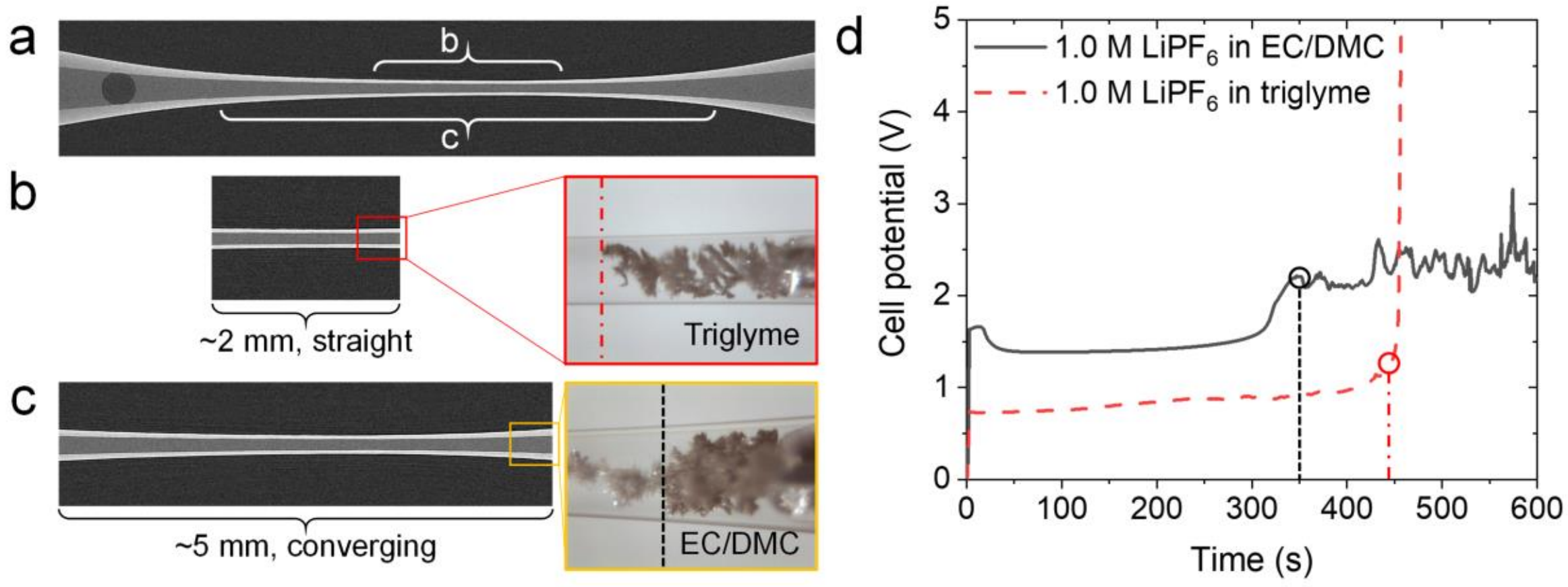

Figure 1. Setup and method of the capillary cell experiments. (a) X-ray computed tomography image of a pulled glass capillary. (b) Capillary cells with triglyme electrolyte enable the shorter interelectrode distance of about 2 mm, where the capillary appears straight. (c) Capillary cells with the mixed ethylene carbonate and dimethyl carbonate (EC/DMC) electrolyte only allow electrodes to reach the tapering part of the capillary, leaving an interelectrode distance about 5 mm. (d) Voltage responses of galvanostatic electrodeposition at over-limiting current densities for the short but straight capillary cell with triglyme electrolyte and the long but converging capillary cell with mixed ethylene carbonate and dimethyl carbonate (EC/DMC) electrolyte. Open circles mark the corresponding Sand’s time for each case.

When determining the Sand's time, both the morphological transitions shown in the optical microscopy images (Fig. 1b, c) and the abrupt changes in the voltage responses (Fig. 1d) were considered. For the EC/DMC system (1M $LiPF_6$ in EC/DMC, 1:1 v/v), when the Sand's time was reached, the voltage jumps abruptly, followed by a fluctuating plateau; and the corresponding optical microscopy image shows the transition from mossy to dendritic growth, similar to our previous report.[5] For the triglyme system, a similar voltage jump was observed, but without the fluctuating plateau. The voltage diverges to the safety limit of the battery tester. The growth, in this case, stopped at the time of the voltage spike. While a clear transition from mossy to dendritic growth was not observed when using this particular solvent, Sand's time can still be accurately determined, which suffices our investigation of the concentration polarization dynamics.

Figure 2 displays the experimental results obtained in cells with converging or straight channels. Since the experiments always end before reaching the center of the symmetrical capillary cells, the growths of lithium only occur in the converging part of the channel. The logarithmic plots of Sand's time versus current density obtained from long cells with clear converging channels always yield slopes greater than -1.5, regardless of the choice of electrolyte (Fig. 2a and 2b), but are consistent with our previous report.[5] In contrast, the slope of the data obtained from short straight cells showed a scaling exponent of -1.94 (Fig. 2c), very close to the theoretical value (-2) predicted by the classic Sand's equation.

The experimental results above suggest that the converging channel results in a positive deviation of the Sand's time scaling, where significant shortening of the Sand's time at low overlimiting current densities was observed. To verify this observation, an axisymmetric 3-dimensional (3D) model of the capillary cells, with either a straight line or the catenary curve as the profile of the inner wall, were established in COMSOL Multiphysics. The effective electrodiffusion equation was numerically solved within the 3D geometries:

$$\frac{\partial c_i}{\partial t} = D_{amb} \nabla \cdot (\nabla c_i), D_{amb} = \frac{D_+ D_- Z_+ - D_+ D_- Z_-}{D_+ Z_+ - D_- Z_-} \quad (1)$$

Here, $c_i$ is the concentration of the species $i$, $D_i$ the diffusion coefficient of species $i$, and $Z_i$ the charge number. Details of the equation and the boundary conditions used in the simulations can be found in the Methods section and the Supplementary Information (S1-11, Table S1). The Li ion concentration distribution was computed until the concentration at the electrode surface reached zero, i.e. Sand's time, as demonstrated in Figure 2d. Here, we also included the advancing electrode front to match our experimental observation. With everything else identical, simply changing the straight channel to a converging channel shortens the Sand's time by 55.4% (from $t_S$ =3440 s to $t_S$=1534 s). Using the widely accepted values[41,42] of diffusion coefficient ($3\times10^{-6}$ $cm^2$ $s^{-1}$) and Li ion transference number (0.38), the Sand's times calculated by the COMSOL models match very well with our experimental results for all cases (Fig. 2a-c).

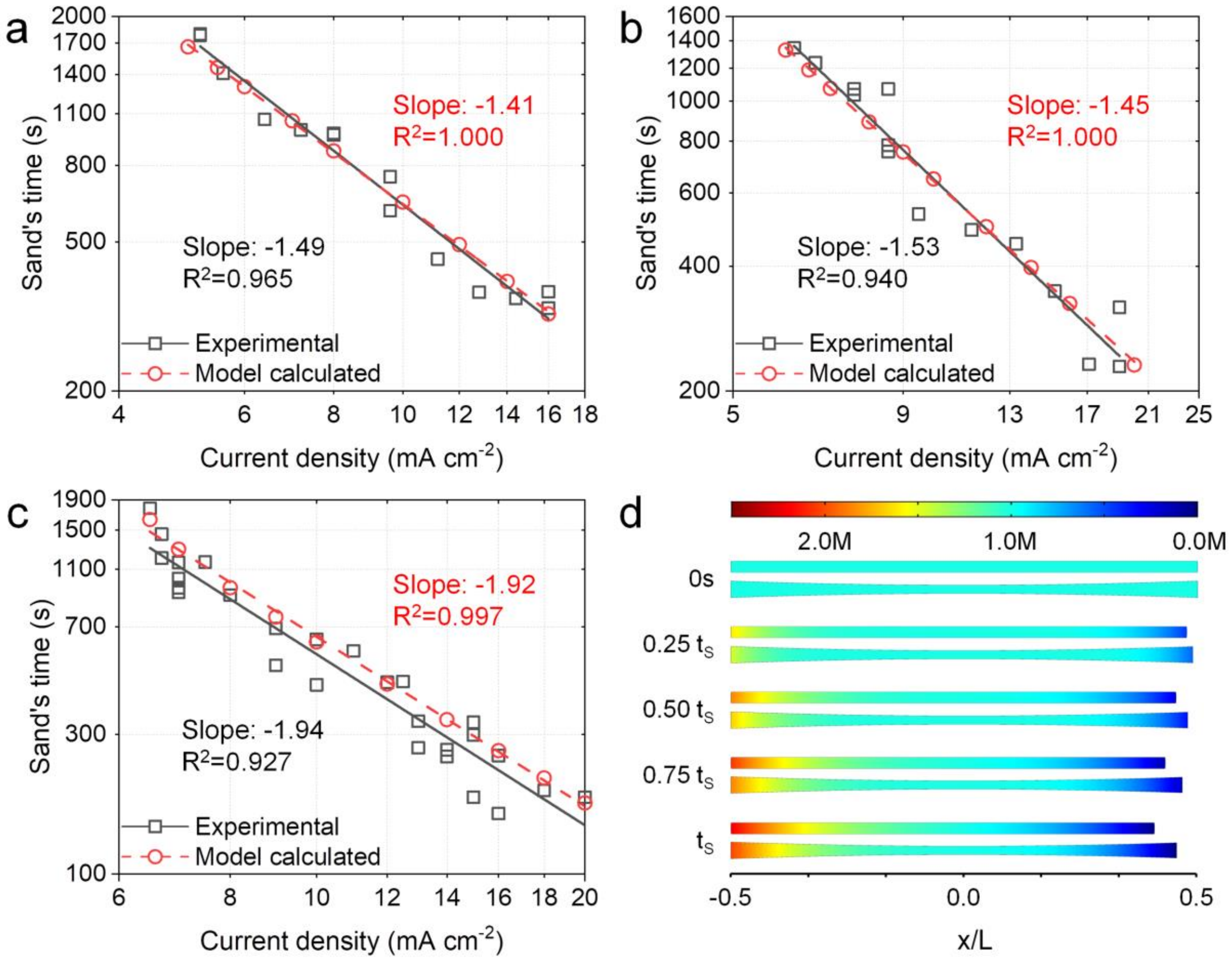

Figure 2. Comparisons between experimental Sand's time versus current density results and mathematical simulations. (a) Case 1: Long capillary cells with a 5.0 mm interelectrode distance and 1.0 M $LiPF_6$-EC/DMC electrolyte. (b) Case 2: Long capillary cells with a 4.5 mm interelectrode distance and 1.0 M $LiPF_6$-triglyme electrolyte. (c) Case 3: Short straight capillary cells with a 2.0 mm interelectrode distance and 1.0 M $LiPF_6$-triglyme. The diffusion coefficient of $3\times10^{-6}$ $cm^2$ $s^{-1}$ was used for all cases. Lines are the best fits for the experimental and simulation data points. (d) Mathematical simulation results of the concentration polarization with the advancing working electrode front for cells with both the straight and converging channels. The total volumes and the interelectrode distances (5 mm) were set equal for both cells.

## 2.3 Effects of the Channel Geometry and the Advancing Electrode Front

Although the comparison between the experimentally measured and simulated Sand's times strongly suggests that the inner wall curvature of the channel and the advancing electrode front nicely explain the deviation of the Sand's time scaling exponent, decoupling these two effects could better elucidate the transport dynamics inside the channels.

As shown in Fig. 3a, Sand's times in converging channels are significantly shorter than that in straight channels even without the advancing electrode. This is true for all overlimiting current densities, but more so for the relatively lower overlimiting current densities, which yields a positive deviation of the slope (from -2 to -1.77). As the concentration gradient extends toward the center of the cell, i.e. the increasingly narrower part of the channel, the absolute number of ions needs to be depleted decreases, making depletion faster than that in straight channels. It's noteworthy that delayed diffusion processes inside the channel with a varying cross-section were considered as a diffusion past an entropy barrier,[43] and have also been studied in terms of a modified Fick-Jacobs equation adopting mathematically convenient effective diffusion coefficients.[43–50] In our simulations, however, no artificial change of the diffusion coefficient was adopted.

Curvature effects alone cannot change the Sand's time properly to match the experimental results. As observed in experiments, we introduced the advancing electrode front based on the charge and mass conservations, via using a metal porosity ($\varepsilon_M$), to match the lengths of observed growths. For straight channels, the advancing electrode front significantly delays the Sand's time (Fig. 3b). While the salt near the electrode surface is being depleted, the advancing electrode front counteracts the depletion process and makes the Sand's time longer. Surprisingly, the effect of the advancing electrode applies uniformly to all current densities, and the resulting Sand's time scaling is not affected. However, for converging channels (Fig. 3c), the same effect impacts the higher overlimiting current densities more significantly while extending the Sand's time for all current densities. The slope, therefore, deviates more positively (from -1.77 to -1.63).

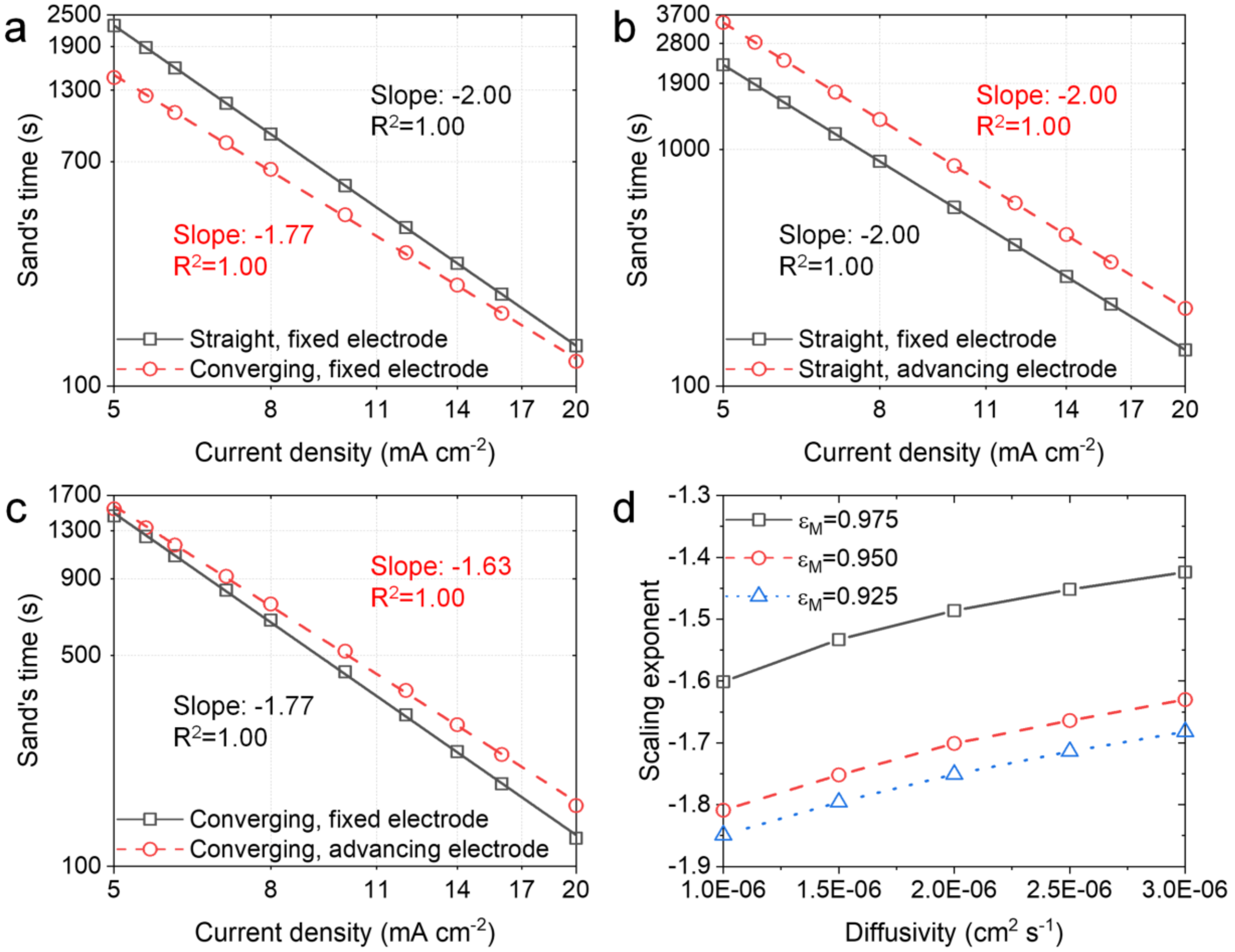

Figure 3. Synergistic effects from curvature and electrode growth. (a) Simulation results of the impact of curvature on the Sand's time scaling with fixed interelectrode distance. (b) Simulation results reveal that the advancing electrode front has no effect on the Sand's time scaling for straight channels. (c) Simulation results show that the advancing electrode fronts in the converging channel

further extend the Sand's time, with a stronger effect for higher current densities. (d) For converging channels, higher metal porosity, i.e. faster electrode growth toward the counter electrode, and higher diffusion coefficient are responsible for larger deviation in Sand's time scaling exponent.

The verified COMSOL model allows versatile analyses that are hard to be realized in experiments. It also offers the opportunity to perform a sensitivity analysis on the key physical parameters. As shown in Fig. 3d, the effects of the diffusion coefficient and the metal porosity ($\varepsilon_M$) for electrodeposition in converging channels were assessed. Both the higher diffusivity and higher metal porosity (sparser structure and longer growths) resulted in a more positively deviated Sand's time scaling exponent, with a slight nonlinearity. Further analyses revealed that the converging channel would lower the limiting current, while the advancing electrode front would counteract the trend, leading to a nonlinear concentration profile at the limiting current density. See more details in the Supplementary Information (Fig. S2, Eq. S12-22).

## 2.4 Analytical approximation of the dynamics

The commonly used diffusion equation, Eq (1), assumes constant cross-sectional areas along the diffusion path. To simplify the dynamics to one-dimensional for possible analytical solutions, Eq (2) that included variable cross-sectional area $A(x)$ is used instead:

$$\frac{\partial c(x,t)}{\partial t} = D_{amb}\frac{\partial^2 c}{\partial x^2} + D_{amb}\frac{dA(x)}{A(x)dx}\frac{\partial c}{\partial x} \tag{2}$$

In general, direct analytical solutions for Eq (2) can hardly be obtained. However, given that the concentration polarization and metal growths we investigated did not extend too deep toward the center of the cell (bulk electrolyte), we can introduce an exponential function $A(x) = A_0 \exp(bx)$ to approximate the varying cross-sectional area starting from the initial position of the electrode surface (Figs. 4a and 4b). Therefore, the factor $(dA(x)/dx)/A(x)$ in Eq (2) reduces to a scalar $b$, which represents the changing rate of the cross-sectional area along the axial direction. The approximation equation resembles the simplest advection-diffusion equation and can be solved analytically to obtain the concentration profile $c(x,t)$. The detailed derivation and solution procedures of Eq (2) are available in the Supplementary Information Eqs (S23-31). By setting the concentration at the electrode surface to zero, i.e. $c(x = 0, t = t_S)$, an implicit formula for $t_S$ as a function of the applied current density $J$ can be obtained:

$$2\sqrt{\frac{t_S}{D_{amb}\pi}} \exp\left(-\frac{b^2 D_{amb} t_S}{4}\right) - b t_S \operatorname{erfc}\left(\frac{b\sqrt{D_{amb} t_S}}{2}\right) = \frac{c_0 n F}{J(1 - t_+)} \tag{3}$$

where erfc is the complementary error function, $J$ the applied overlimiting current density. The geometry of the channel, as reflected by the area changing rate $b$, indeed affects both terms on the left-hand side.

As revealed in the X-ray CT image of our pulled capillary cells, different interelectrode distances within the pulled glass capillary result in different curvatures near the electrode. Therefore, in our simulations, we truncated the catenary curve and used the exponential function $A(x)$ to obtain the fitted area changing rate $b$, which then can be used to obtain the analytical Sand's time from Eq (3). On the other hand, COMSOL simulations with the truncated catenary curve were performed to get the simulated Sand's time. As shown in Fig. 4, our implicit analytical formula for Sand's time matches very well with the simulation results, which was ensured by the good fitting results of the exponential function (Figs. 4a and 4b). Even for the relatively straight cell of 2 mm in length, the exponential function can accurately capture the trend up to 1.5 times the estimated maximum diffusion length ($\sqrt{D_{amb}t_{max}} \approx 0.7$ mm, where $D_{amb}$ is the ambipolar diffusivity and $t_{max}$ is maximum time considered in the COMSOL simulations). With the fitted area changing rate *b*, the analytical Sand's time versus current density for each of the selected cells was plotted in Fig. 4c to find out the effective analytical scaling exponent. The scaling exponent, unexpectedly, varies non-monotonically as the interelectrode distance increases from 2 mm to 10 mm, with a maximum value near 5.5 mm (Fig. 4d). The non-monotonic trend is originated in the non-monotonic variation of changing rate *b* versus the interelectrode distances. As shown in Figure S3, the 5.5-mm-long cell exhibit the most significant area changing rate. More generally, however, an effective Sand's time scaling exponent ($n$ in $t_S \sim J^n$) as a function of the area changing rate $b$ can be obtained:

$$n = -0.0006b^2 - 0.0328b - 2 \quad (4)$$

The noticeable differences for cells with smaller interelectrode distances (Fig. 4d) are likely caused by the inaccuracy of $A(x)$ approximations. Including the advancing electrode front in the simulations leads to more positive deviation of the scaling exponents for all cases, while the nonmonotonic trend is preserved.

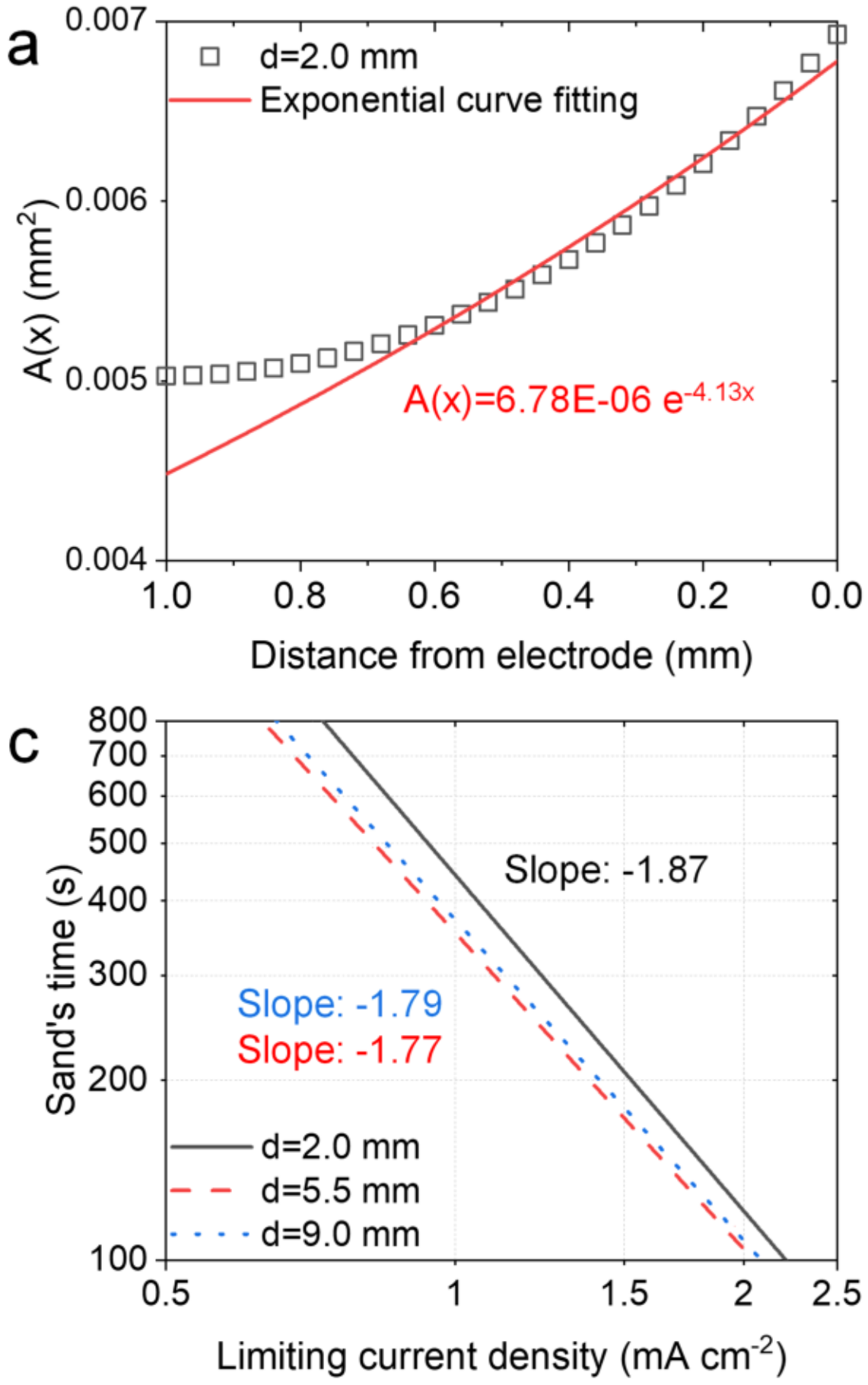

Figure 4. Analytical approximation of the Sand's time scaling dependence on the interelectrode distance. (a,b) Analytical approximation of the cross-sectional area of the cells with different interelectrode distances. (c) Analytical results of Sand's behavior for the three cases. (d) Comparisons between the simulated and the analytical Sand's time scalings for fixed electrode and advancing electrode cases.

## 2.5 Sand's Time in Expanding Channels

We have demonstrated how the converging channel and the advancing electrode front affect the concentration polarization dynamics, as reflected by the absolute values and the scaling exponent of Sand's time with respect to various overlimiting current densities. Here, we extend our analyses for expanding channels. Simulations were performed in COMSOL with the same equations and boundary conditions but with different geometry. To better compare with the symmetrical pulled glass capillary, an artificial spindle-like symmetrical geometry with expanding channels toward the center was adopted (Supplementary information, Fig. S4a). In our simulations, the growth of the metal electrode did not reach the center and stayed only in the expanding part. Apparently, opposite to the scenario in converging channels, the increasing cross-sectional area in expanding channels provides extra ions to the electrode surface as it grows toward the center of the cell, which delays or even avoids the complete concentration depletion at the advancing electrode front. As a result, both the system-specific limiting current density and the Sand's time are increased. Such effects appear more significant at relatively low overlimiting current densities, leading to the negatively deviated Sand's time scaling exponent (Fig. 5a). By varying the interelectrode distance of the spindle-shaped cell, we effectively varied the wall curvature near the electrode (Fig. 5b) as we did for converging channels in Fig. 4. Here, shorter interelectrode distance represents a faster area changing rate (scalar $b$) near the electrode, which resulted in more negatively deviated scaling exponents. The advancing electrode front further enhanced this effect (Fig. S4b). Compared with the pulled capillary cell with converging channels toward the center, the steady-state concentration profile for expanding channels was also nonlinear but was curved in the opposite direction (Fig. S4c).

The theoretical predictions on the concentration polarization and Sand's time scaling are confirmed by experimental results. As shown in the inset of Figure 5c, we managed to push the working electrode to the center of the pulled capillary cell, such that the channel from the working electrode (right) to the counter electrode (left) has an expanding geometry. As expected from the simulation results, the ion depletion inside expanding channels would be delayed and the limiting current density values significantly increased. Depletion was indeed observed at much higher current densities ( > 40 mA $cm^{-2}$.) The galvanostatic charge curve (Fig. 5c) shows that the cell potential starts high, then decreases as the electrode front advances toward the bulk, which can be attributed to the increasing cross-sectional area of the channel and therefore the decreasing actual current density. Consistent with earlier results using the triglyme-based electrolyte, the complete concentration depletion can be inferred from the sudden increase of cell potential and the termination of the growth process. The Sand's times thus determined were displayed in the logarithmic plot against the current densities, which yielded a scaling exponent close to -4.19 (Fig. 5d). This negative deviation from the exponent of -2 for ideally straight channels confirms our theoretical prediction qualitatively. The noticeable quantitative difference between the simulations and the experiments can be attributed to the nonlinear metal growth speed (Fig. S5). At such high current densities and therefore high overpotentials, the electroosmotic instability and convection may have played a more significant role than in the converging and straight channels where much lower current densities were applied. The most important implication of the results, however, is that expanding electrolyte channels, if can be engineered, can actually alleviate or even avoid the concentration depletion, such that metal penetration by fast-advancing tip-growing dendrites can be avoided.

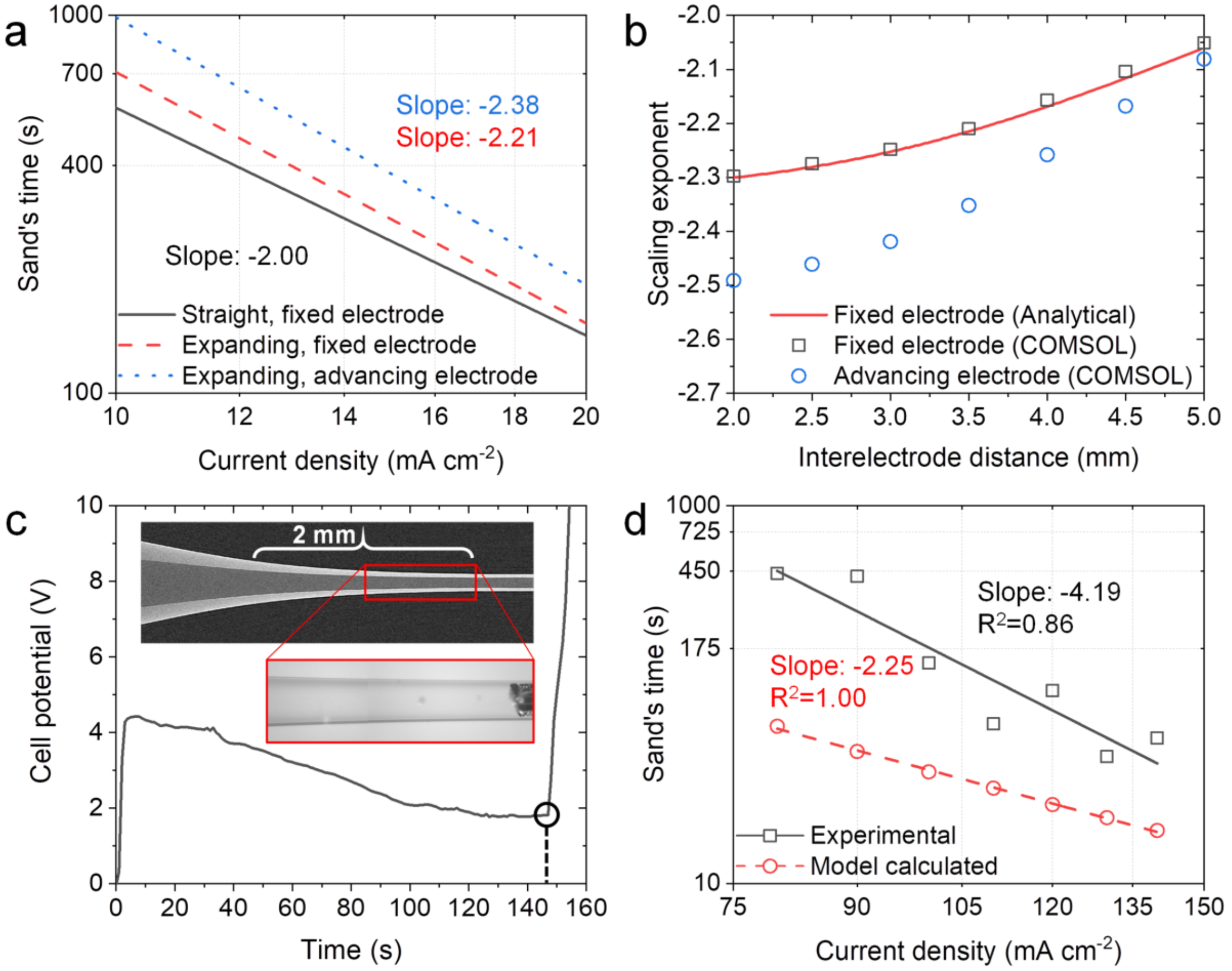

Figure 5. Theoretical and experimental analyses on the concentration polarization in expanding channels. (a) Comparison of simulation results inside various channel configurations. (d=3.5 mm) (b) Comparisons between the simulated and analytical Sand's time scaling for fixed electrode and advancing electrode cases in expanding channels. (c) Cell configuration and example galvanostatic charge curve for the expanding channel. The electrolyte used was 1.0M $LiPF_6$ in triglyme and the current density was 100 mA $cm^{-2}$. The open circle marks the Sand's time. (d) Logarithmic plot of experimental and simulated Sand's times versus current densities obtained within expanding channels. For the simulations of the channels with advancing electrode front, the metal deposit porosity of 0.9 was used for (a) and (b), and 0.978 for (d).

# 3 Discussions

## 3.1 Macroscopic converging areas through porous separators

While our pulled capillary cells offer a well-defined system to investigate the subtleties of the localized transport dynamics in a single channel, penetration in practical batteries would occur in multiple channels. Interestingly, experimental results in the literature showed that the collective cross-sectional area of all penetration channels converges along the thickness of the separator/electrolyte.[51] On the other hand, positively deviated Sand's time scaling exponent between the cumulative time before penetration and the cycling current density applied to practical coin cells was repeatedly reported. Choudhury *et al.*[52] reported a value of -1.84 and -1.35 for a crosslinked nanoparticle-polymer composite electrolyte and a porous separator, respectively. Lu *et al.* reported several deviated values, often less than but close to -1, obtained from multiple systems with separators: ionic liquid-nanoparticle hybrid system,[53,54] salt-reinforced liquid system,[55] and a liquid system with AAO separator.[56] While the physical scenario for concentration polarization during battery cycling is different from that for one-way polarization, the positive deviation of the scaling exponent and the converging penetration area toward the counter electrode coincide with, and may find the ultimate explanations from the insights of our analyses.

## 3.2 Penetration paths in solid-polymer electrolytes

In contrast, polymer electrolytes without inert porous structures were found to yield scaling exponents close to -2 between the cumulative time before penetration and the cycling current density. [57–60] The exact value of -2 was reported by Schaefer *et al.*[61] from cells using a freestanding gel-type electrolyte, but with $SiO_2$ nanoparticles. The volume fraction of the nanoparticles was controlled to be 0.15, much lower than the inert volume fraction (0.6) of typical porous separators with porosity around 0.4. According to our results, this nearly perfect scaling exponent suggests that the collective area of all penetration channels remained constant along the thickness of the electrolyte layer, which indeed has been revealed by *in situ* characterization methods.[21,62] According to our simulations, the Sand's time scaling exponent for isolated straight channels is not affected by the advancement of the electrode surface (Fig. 3b), but the absolute values of the Sand's time are significantly extended. Interestingly, earlier reports by both Khurana *et al.*[59] and Huang *et al.*[60] demonstrated the significant extensions of Sand's time without the deviation of the scaling exponent. Still, the positive deviation of the Sand's time scaling found in similar polymer electrolyte systems may imply a converging collective area of penetration channels, likely due to the percolated inert nanoparticles or simply less-conductive domains.[24–26]

Our combined experimental and theoretical analyses offer a clear explanation to the question pointed out by Rosso *et al*.[57]: why an underlimiting current would induce dendrite penetrations, which further yield a scaling of -2 between penetration time and the current density, to resemble the Sand's time scaling for diffusion-limited dendrite initiation. Initially, when a small current is applied to an equilibrium system, an underlimiting current density that distributed over the entire surface of the flat electrode is achieved. As the system is being polarized, local physical heterogeneities and dynamic instabilities start to divert the flux onto just a few hot spots. When the newly formed metal electrode front enters a pore of the separator or the solid electrolytes, the localized electric field amplifies the flux at the metal electrode tip even further, which would easily exceed the limiting current density. When this naturally evolved local overlimiting current density start to polarize the concentration within that microchannel or conduction path, it will start to follow the classic or our new implicit Sand's formula to reach diffusion limitation. Once a dendritic growth is triggered, the system will be short-circuited almost instantly. The penetration time is dominated by the classic or modified Sand's time for concentration depletion, as Schaefer et. al. also observed in their experiments.[61]

# 4 Conclusions

Motivated by the "black spots" observed in short-circuited batteries, we used our special glass capillary cells to investigate the concentration polarization and metal dendrite penetration through these electrolyte channels. Our combined experimental and theoretical analyses, together with the examinations of literature data, showed that an underlimiting current density applied to the entire surface of the electrode can evolve into an overlimiting current density localized onto just a few "black spots". Within the penetration channels that connecting the black spots on both sides of the separator or electrolyte, the time to trigger metal dendrites scales with the actual local current density via $t_S \sim J^n$. For straight channels or curved channels with a constant cross-sectional area, the scaling exponent equals -2, as predicted by the classic Sand's equation. For converging channels, i.e. with decreasing cross-sectional area, the scaling exponent exhibits a positive deviation. For expanding channels, i.e. with an increasing cross-sectional area, the scaling exponent shows a negative deviation. Our results stress that while batteries are designed to avoid reaching the limiting current density estimated by using the geometrical area of the electrode/electrolyte, the structural heterogeneities and dynamic instability can lead to highly localized current density. Such a local current density easily exceeds the limiting current density to trigger the diffusion-limited, tip-growing, yet fast-advancing metal dendrites through the isolated electrolyte channels. It is therefore recommended to obtain the Sand's time scaling to assess the penetration structure through the separator or electrolyte layers, so that more accurate penetration time, rather than the ideal Sand’s time, can be obtained for safety designs. Engineering homogeneous electrolyte, or electrolyte with expanding conduction structures are favorable options for safety improvements.

# 5 Methods

## 5.1 Glass capillary cell experiment

Electrochemical cells with the 2-electrode configuration were fabricated in the pulled glass capillaries. Two electrolyte systems were used: 1.0 M $LiPF_6$ in EC/DMC, 1:1 v/v (battery grade, Sigma-Aldrich); and 1.0 M $LiPF_6$ in triglyme ($LiPF_6$ salt, battery-grade > 99.99%; triglyme, 99.5%). For the triglyme system, both components were purchased from Sigma-Aldrich and mixed and stirred for 12 hours before use. One major difference is that the lubricity of triglyme is higher than that of the mixed EC/DMC solvent, which allows the lithium metal electrodes to be pushed further inside the capillary, achieving a shorter distance between them. Since the profile of the interior capillary wall is close to a catenary curve, it closely approximates a straight line as the distance between the electrodes decreases. Three experimental cases were compared to analyze the converging channels: (1) 1.0 M $LiPF_6$ in triglyme, with an interelectrode distance, d, of $\approx$ 2 mm. (2) 1.0 M $LiPF_6$ in triglyme, d $\approx$ 4.5 mm. (3) 1.0 M $LiPF_6$ in triglyme, d $\approx$ 5 mm. Cases (1) and (3) were compared to analyze the effect of the wall curvature, while (1) and (2) were compared to prove that the effect was not a result of different solvents in the system. For the analysis on the expanding channels, slightly modified capillary cells, with d $\approx$ 2 mm but with shifted configuration was used, so that only the expanding part of the capillary would be included in the system. All the cell assembly and electrochemical experiments were done inside a glovebox filled with argon gas with water and oxygen concentration less than 0.5 ppm. Electrochemical testing was done with a Gamry potentiostat (Reference 600+, Gamry Instruments). *Operando* imaging was done with an optical microscope (MU500, AmScope).

## 5.2 COMSOL simulation

COMSOL Multiphysics was used to simulate the transport phenomena inside the capillary cells. More details on the equation and the parameters used in the COMSOL model can be found in the Supplementary Information (S1-11, Table S1).

A longitudinal cross-section image of the pulled capillary was produced with a micro-CT X-ray scan (Zeiss Xradia Versa 520 XRM), and the image was digitized into data points that were fitted with a catenary curve. The fitted curve was used in the COMSOL model when defining the axisymmetric geometry.

## Acknowledgements

This work is supported by a National Science Foundation grant (Award No. 1934122). The materials characterization experiments were partially supported by IMSE (Institute of Materials Science and Engineering) and by a grant from InCEES (International Center for Energy, Environment and Sustainability) at Washington University in Saint Louis. P.B. acknowledges the startup support from Washington University in St. Louis.